\documentclass{article}
\usepackage{hello}

\usepackage[margin=1in]{geometry}
\usepackage{graphicx}
\usepackage{booktabs}
\usepackage{array}
\usepackage{amsmath}
\usepackage{bbm}
\usepackage{amssymb}
\usepackage{amsfonts}
\usepackage{multirow}
\usepackage{verbatim}
\usepackage{graphicx}
\usepackage{caption}
\usepackage{url}
\usepackage{longtable}
\usepackage{supertabular}
\usepackage{float}
\usepackage{enumitem}
\usepackage{tablefootnote}
\usepackage[round,semicolon]{natbib}
\usepackage{xcolor}
\usepackage{colortbl}
\usepackage{xspace}
\usepackage{textcomp}
\usepackage{makecell}
\usepackage{multirow}
\usepackage{lscape} 
\usepackage{siunitx}
\usepackage{tabularx}
\usepackage{comment}

\usepackage{amssymb}
\usepackage{pifont}
\usepackage{scrextend}
\usepackage{float}
\usepackage{array}
\usepackage{tgpagella}
\usepackage{latexsym}
\usepackage[T1]{fontenc}
\usepackage[utf8]{inputenc}
\usepackage{microtype}
\definecolor{mydarkblue}{rgb}{0,0.08,0.45}
\usepackage[colorlinks,citecolor=mydarkblue,urlcolor=mydarkblue,linkcolor=mydarkblue]{hyperref}
\usepackage{url}            
\usepackage{nicefrac}       
\usepackage{changepage}
\usepackage{xargs}          
\usepackage{wrapfig,lipsum,booktabs}
\usepackage{longtable}
\usepackage{caption}
\usepackage{subcaption}
\usepackage{endnotes}
\usepackage{tablefootnote} 

\usepackage{pgfplots}
\usetikzlibrary{pgfplots.groupplots}
\pgfplotsset{compat=1.3}
\usepackage{tikz}
\usetikzlibrary{patterns}
\usepackage[most]{tcolorbox}
\usepackage{CJKutf8}

\usepackage[capitalize,noabbrev]{cleveref}
\crefname{section}{Section}{\S\S}
\Crefname{section}{Section}{\S\S}
\crefname{table}{Table}{Tables}
\crefname{figure}{Figure}{Figures}
\crefname{algorithm}{Algorithm}{}
\crefname{equation}{eq.}{}
\crefname{appendix}{Appendix}{}
\crefformat{section}{Section #2#1#3}
\usepackage{multicol}
\usepackage{multirow}
\usepackage{bm}

\usepackage{tcolorbox}
\usepackage{titlesec}
\titleformat*{\section}{\large\bfseries}

\definecolor{battleshipgrey}{rgb}{0.3, 0.3, 0.3}
\definecolor{brilliantrose}{rgb}{1.0, 0.33, 0.64}
\definecolor{americanrose}{rgb}{1.0, 0.01, 0.24}
\definecolor{jweigreen}{rgb}{0,0.45,0.24}
\definecolor{bluegray}{rgb}{0.1, 0.1, 0.4}
\definecolor{ao(english)}{rgb}{0.0, 0.5, 0.0}
\definecolor{blanchedalmond}{rgb}{1.0, 0.92, 0.8}
\definecolor{atomictangerine}{rgb}{1.0, 0.6, 0.4}
\definecolor{chocolate(web)}{rgb}{0.82, 0.41, 0.12}
\definecolor{bananayellow}{rgb}{1.0, 0.88, 0.21}
\definecolor{goldenbrown}{rgb}{0.6, 0.4, 0.08}
\definecolor{aliceblue}{rgb}{0.94, 0.97, 1.0}
\definecolor{beige}{rgb}{0.96, 0.96, 0.86}
\definecolor{babyblue}{rgb}{0.54, 0.81, 0.94}
\definecolor{camel}{rgb}{0.76, 0.6, 0.42}
\definecolor{cinnamon}{rgb}{0.82, 0.41, 0.12}
\definecolor{deepskyblue}{rgb}{0.0, 0.75, 1.0}
\definecolor{frenchblue}{rgb}{0.0, 0.45, 0.73}
\definecolor{classicrose}{rgb}{0.98, 0.8, 0.91}
\definecolor{frenchrose}{rgb}{0.96, 0.29, 0.54}
\definecolor{frenchlilac}{rgb}{0.53, 0.38, 0.56}
\definecolor{frenchbeige}{rgb}{0.65, 0.48, 0.36}
\definecolor{verylightgreen}{RGB}{240, 255, 235}
\definecolor{verylightred}{RGB}{255, 235, 235}
\definecolor{verylightyellow}{RGB}{255, 254, 235}
\definecolor{dt}{gray}{0.7}

\definecolor{forestgreen}{HTML}{2e7d43}
\definecolor{color1}{HTML}{FF9999}
\definecolor{color2}{HTML}{FF6666}
\definecolor{color3}{HTML}{FF3333}
\definecolor{color4}{HTML}{E60000}
\definecolor{color5}{HTML}{B30000}
\definecolor{color6}{HTML}{8CD98C}
\definecolor{color7}{HTML}{53c653}
\definecolor{color8}{HTML}{39ac39}
\definecolor{color9}{HTML}{2d862d}
\definecolor{color10}{HTML}{206020}
\definecolor{color11}{HTML}{cca300}

\usepackage{minitoc}

\title{
\textbf{
OSUM: Advancing Open Speech Understanding Models with Limited Resources in Academia}
}

\author{
\large{}
Audio, Speech and Language Processing Group (ASLP@NPU)\\
School of Computer Science, Northwestern Polytechnical University \\
\small{}
\url{http://www.npu-aslp.org} \\
\small{}
Code \& Demo \& Models: \ \ \url{https://github.com/ASLP-lab/OSUM}
\thanks{This is the v2 version of the technical report. The experimental results reported herein differ from those in v1 because of adding new data and training in more steps.}
}

\date{}

\begin{document}
\begin{CJK}{UTF8}{gbsn}

\doparttoc 
\faketableofcontents 

\maketitle

\begin{abstract}
\noindent
Large Language Models (LLMs) have made significant progress in various downstream tasks, inspiring the development of Speech Understanding Language Models (SULMs) to enable comprehensive speech-based interactions.
However, most advanced SULMs are developed by the industry, leveraging large-scale datasets and computational resources that are not readily available to the academic community.
Moreover, the lack of transparency in training details creates additional barriers to further innovation.
In this study, we present OSUM, an Open Speech Understanding Model designed to explore the potential of training SLUMs under constrained academic resources.
The OSUM model combines a Whisper encoder with a Qwen2 LLM and supports a wide range of speech tasks, including speech recognition (ASR), speech recognition with timestamps (SRWT), vocal event detection (VED), speech emotion recognition (SER), speaking style recognition (SSR), speaker gender classification (SGC), speaker age prediction (SAP), and speech-to-text chat (STTC).
By employing an ASR+X training strategy, OSUM achieves efficient and stable multi-task training by simultaneously optimizing ASR alongside target tasks.
Beyond delivering strong performance, OSUM emphasizes transparency by providing openly available data preparation and training methodologies, offering valuable insights and practical guidance for the academic community.
By doing so, we aim to accelerate research and innovation in advanced SULM technologies.
\end{abstract}

\section{Introduction}
\label{sec:intro}

\begin{figure*}[ht!]
\centering
\includegraphics[width=12.5cm]{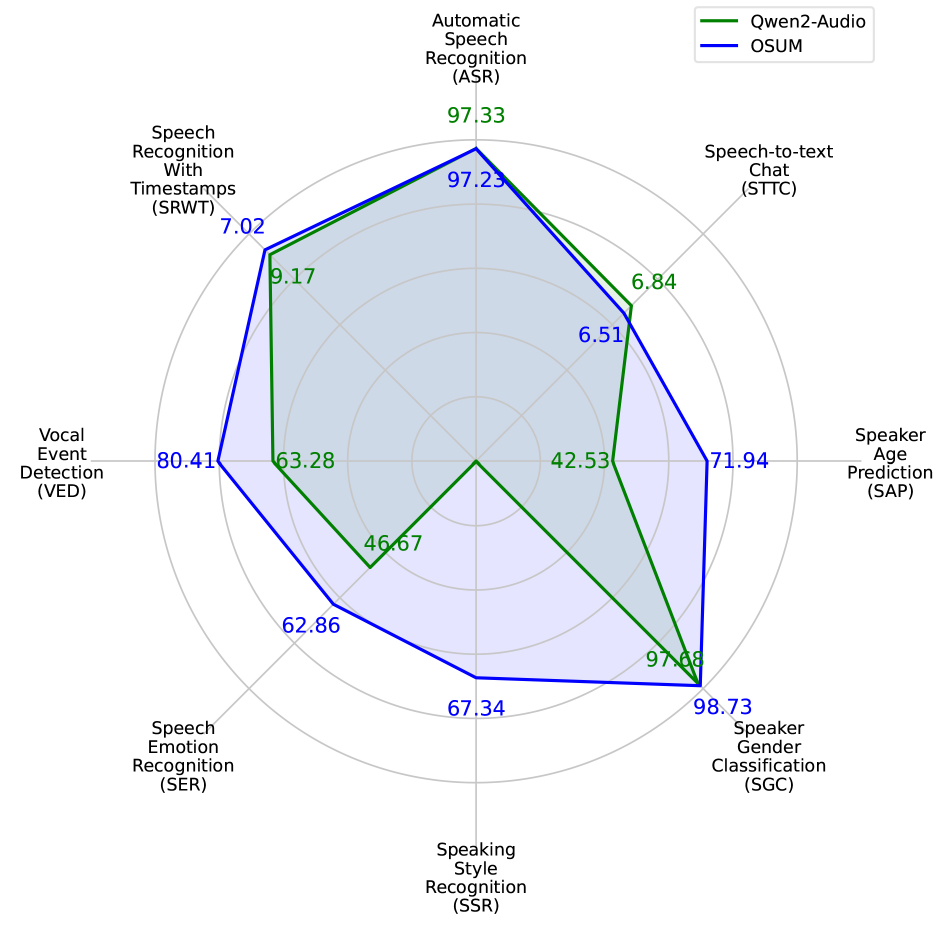}
\caption{Comparison of Qwen2-Audio and our OSUM model. In most tasks, OSUM achieves a better performance than Qwen2-Audio despite using significantly fewer computational resources and training data.
The values for each model's task in the radar chart are based on the average results on the public and internal test sets, as shown in Table~\ref{tab:res_asr} and Table~\ref{tab:res_multi}.
}
\label{fig:radar}
\end{figure*}

Large language models (LLMs) have shown tremendous progress towards Artificial General Intelligence (AGI) in recent years. 
Given the inherent human preference for speech-based interaction, there has been growing interest in extending LLMs with speech capabilities to develop Speech LLMs.
To generate fluent and expressive text or speech responses, Speech LLMs must fully comprehend input speech, including both its semantic content and paralinguistic information, like emotion, speaking style, speaker gender, and age.
Moreover, this comprehension ability is also crucial for audio data labeling. Currently, the mainstream multi-label generation approach is to use multiple models to label each task separately, which consumes extremely high computational resources. A labeling model capable of accurately generating multiple labels simultaneously holds broad application prospects. 

The area which focuses on Speech Understanding Language Models (SULMs), has seen notable advancements through projects such as Qwen-Audio\citep{chu2023qwen_audio}, Qwen2-Audio\citep{chu2024qwen2_audio}, PandGPT\citep{su2023pandagpt}, and SALMONN~\citep{tangsalmonn}.
Whisper~\citep{radford2023whisper} marks a pioneering exploration of speech understanding independent of LLMs, utilizing an encoder-decoder Transformer~\citep{vaswani2017attention} architecture to tackle a variety of speech tasks, such as automatic speech recognition (ASR), speech translation (ST), language identification (LID), and voice activity detection (VAD).
Building on Whisper’s design, SenseVoice~\citep{an2024funaudiollm} and TouchASP~\citep{song2024touchasp} expand more tasks like speech emotion recognition (SER) and audio event detection (AED), further enriching their ability to process and comprehend human speech.
Qwen-Audio integrates Whisper's encoder with the text-based Qwen LLM~\citep{bai2023qwen}, enabling the latter to understand speech. Compared to Whisper, Qwen-Audio leverages a more powerful LLM decoder and performs over 30 speech-related tasks, making it a representative model in the field of SULMs.
Its successor, Qwen2-Audio, further enhances these capabilities by supporting natural language prompts and achieving superior performance across various benchmarks~\citep{chu2024qwen2_audio}.

Although these advanced SULMs have achieved remarkable progress, most of them are developed by industry, leveraging millions of hours of training data and massive GPU resources.
For instance, TouchASP and SenseVoice utilized 1,000,000 and 400,000 hours of training data, respectively.
Such large-scale resources are typically beyond the reach of academia institutions.
Furthermore, while inference models are often open-sourced, essential details regarding data preparation, training strategies, codebases, and hyper-parameters configurations are rarely disclosed.
These limitations hinder academic community efforts to further optimize and expand SULM research.
Recently, a growing movement advocating for open science in Speech LLM research has emerged. This movement emphasizes the importance of releasing comprehensive training frameworks, datasets, and methodological details to promote research and innovation.
A notable example is the Open Whisper-style Speech Model (OWSM) series~\citep{peng2023reproducing}, which replicates Whisper-style training using open-sourced tools and publicly available data, significantly advancing public understanding and research on speech understanding models.

In this study, we aim to foster broader academic exploration of SULMs with limited resource demands, encouraging wider research community participation.
To this end, we introduce OSUM, an open SULM with its data processing pipeline and training details publicly available.
The OSUM model integrates a Whisper speech encoder, fine-tuned on a multi-task dataset, with a Qwen2 LLM.
It is capable of performing a wide range of speech tasks, including automatic speech recognition (ASR), speech recognition with timestamps (SRWT), vocal event detection (VED), 
speech emotion recognition (SER), speaking style recognition (SSR), speaker gender classification (SGC), speaker age prediction (SAP), and speech-to-text chat (STTC).
Notably, SSR is a distinctive feature of our OSUM model and serves as a vital component of speech understanding. It enhances the model’s capability by improving contextual comprehension and boosting performance across various downstream speech tasks. Furthermore, it establishes a foundation for enabling more natural and context-aware speech-based interactions.
We adopt an ASR+X training strategy to enhance training stability and reduce resource consumption for our SLUM model, wherein an auxiliary ASR task is optimized alongside the primary target task (denoted as ``X''). 
For instance, during the training of the SER task, we concurrently train the ASR task (ASR+SER) by predicting both transcription and emotion labels for each speech sample.
This multi-task training accelerates modality alignment, enabling the LLM to effectively utilize both textual and acoustic modalities.
Our OSUM model utilizes only 50,500 hours of training data and achieves comparable or superior performance to other SULMs.
The overall performance of OSUM is illustrated in Fig.~\ref{fig:radar}.
The model is trained on Nvidia A6000 GPUs and Huawei Ascend NPUs, supporting inference on both platforms.
The goal of this study is to foster transparency and accelerate progress in the field of SULMs by providing accessible tools and resources for the broader research community.

\section{Methodology}
\label{sec:method}

\begin{figure*}[t!]
\centering
\includegraphics[width=16.5cm]{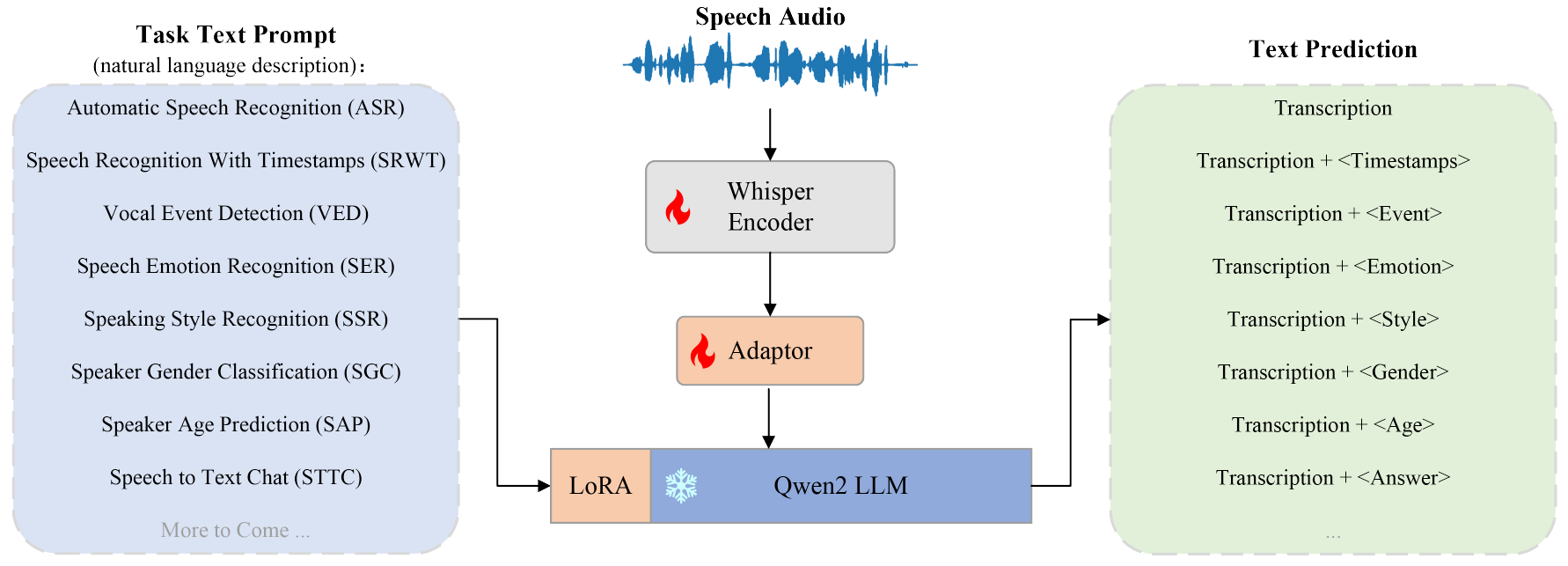}
   \caption{The overview of the architecture and tasks of OSUM.}
\label{fig:framework}
\end{figure*}

This section introduces our proposed OSUM, a model designed for comprehensive speech understanding. 
Section~\ref{sec:model_architecture} presents its architecture;.Section~\ref{sec:multitask_supervised_training} details its multitask training process.  Section~\ref{sec:training_data} and Section~\ref{sec:data_proccessing} provide an overview of the training data and processing pipeline, respectively.

\subsection{Model Architecture}
\label{sec:model_architecture}
As shown in Figure~\ref{fig:framework}, our OSUM model comprises a speech encoder, an adaptor, and an LLM.
During the training, all of the parameters in the encoder and adaptor are updated, while the LLM is fine-tuned with LoRA~\citep{hulora}.
The input of our model consists of a speech and a natural language prompt.
Unlink Whisper~\citep{radford2023whisper} and Qwen-audio~\citep{bai2023qwen}, which rely on instruct tags, the OSUM employs descriptive text, converting all eight supported tasks as shown in Fig.~\ref{fig:framework}. 
Currently, our model supports only text-based responses, but audio output capabilities are under active development.
The following sections describe each sub-module in detail.

\paragraph{Speech Encoder}
Our OSUM utilizes the Whisper-Medium~\footnote{\url{https://huggingface.co/openai/whisper-medium}} model as its speech encoder, which consists of 2 one-dimensional convolutional layers with 2 times downsampling, and 24 Transformer layers with 1024 hidden state dimensions and 16-headed self-attention.
The encoder has approximately 300 million parameters, which makes it take into account both speech comprehension ability and inference efficiency.

\paragraph{Adaptor}
The adaptor module features a hybrid architecture combining 3-layer 1D convolutional layers (Conv1D) and 4-layer Transformer.
The Conv1D layers use kernel widths of (3, 3, 3) and strides of (1, 2, 2), achieving an overall 4 times downsampling.
The Transformer layers have a model dimension of 1,280, an inner dimension of 2,560, and 4 attention heads.
This architecture bridges the output of the speech encoder with the input requirements of the LLM, enabling efficient modality alignment.

\paragraph{LLM with LoRA}
The Qwen2-7B-Instruct is selected as our LLM. Qwen2-7B-Instruct~\footnote{\url{https://huggingface.co/Qwen/Qwen2-7B-Instruct}} is a general-purpose LLM with a parameter scale of 7 billion, specifically designed for multi-task instruction optimization. In our work, we fine-tune the Qwen2-7B-Instruct model using LoRA (Low-Rank Adaptation) technology. The LoRA hyperparameters-${\alpha}$, rank, and dropout ratio are set to 32, 8, and 0.1, respectively.

\subsection{Multitask Supervised Training}
\label{sec:multitask_supervised_training}
The training procedure includes two stages.
First, we perform multi-task supervised fine-tuning on the original Whisper model without an LLM.
Second, we integrate the fine-tuned Whisper encoder with the Qwen2 LLM to create the complete OSUM system, then conduct further supervised training using a larger dataset.

\begin{table}[htp]
\centering
\footnotesize
\caption{Example of prompts and ASR+X labels in OSUM multitask training.}
\label{tab:prompt_label}
\begin{tabular}{@{}clcc@{}}
\toprule
\textbf{Task} & \multicolumn{1}{c}{\textbf{Input}} & \textbf{Prompt} & \textbf{Label} \\ \midrule
\textbf{ASR} & 
\multirow{21}{*}{
\begin{minipage}[b]{0.1\columnwidth}
    \centering
    \raisebox{-.5\height}{\includegraphics[width=\linewidth]{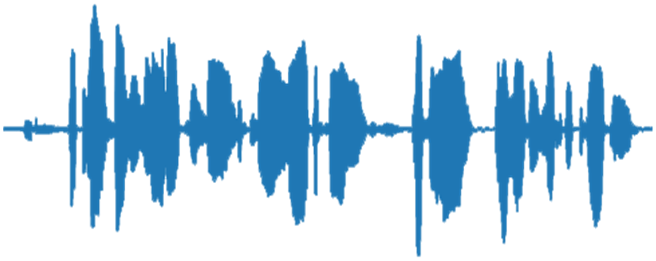}}
\end{minipage}} 
& 转录这段音频中的语音内容为文字。 & 播放小梦想大梦想 \\ \cmidrule(r){1-1} \cmidrule(l){3-4}
\textbf{SRWT} &  & \begin{tabular}[c]{@{}c@{}}请将以下音频进行转录，同时标记出每个英文\\单词及对应中文字符的起始与结束时间，时间\\单位精确到0.1秒，并用<>来表示这些时间范围。\end{tabular} & \begin{tabular}[c]{@{}c@{}}<0.21>父<0.47> <0.46>母<0.60> <0.60>的\\<0.73><0.72>坏<0.86> <0.86>话<1.09>\end{tabular} \\ \cmidrule(r){1-1} \cmidrule(l){3-4} 
\textbf{VED} &  & \begin{tabular}[c]{@{}c@{}}请将以下音频进行转录，并在结尾处给出<音频\\事件>标签，音频事件分为8类：laugh，\\cough，cry，screaming，sigh，\\throat clearing，sneeze，other。\end{tabular} & \begin{tabular}[c]{@{}c@{}}我的属性为了完成我的使命everyday\\ 每个流血不止<SCREAMING>\end{tabular} \\ \cmidrule(r){1-1} \cmidrule(l){3-4} 
\textbf{SER} &  & \begin{tabular}[c]{@{}c@{}}请将以下音频内容进行转录，并在结尾处给出\\<情感>标签，情感共分为8类：sad，anger，neutral，\\happy，surprise，fear，disgust，以及 other。\end{tabular} & \begin{tabular}[c]{@{}c@{}}你一个享受者你有什么跟这个复仇\\ 者去叫板呢<ANGER>\end{tabular} \\ \cmidrule(r){1-1} \cmidrule(l){3-4} 
\textbf{SSR} &  & \begin{tabular}[c]{@{}c@{}}请将以下音频内容进行转录，并在结尾处给出<风格>\\标签，风格共分为8类：新闻科普，恐怖故事，童话\\故事，客服，诗歌散文，有声书，日常口语，其他。\end{tabular} & \begin{tabular}[c]{@{}c@{}}小猪皮革恍然大悟赶紧把这颗白色\\ 的小药片吞进肚子里<童话故事>\end{tabular} \\ \cmidrule(r){1-1} \cmidrule(l){3-4} 
\textbf{SGC} &  & \begin{tabular}[c]{@{}c@{}}请将这段音频进行转录，并在文本末尾附加<性别>\\标签。性别分为两种：female，male。\end{tabular} &  帮我调大声音<MALE> \\ \cmidrule(r){1-1} \cmidrule(l){3-4} 
\textbf{SAP} &  & \begin{tabular}[c]{@{}c@{}}请将以下音频内容转录成文字，并在文字的最后\\加上<年龄>标签，标明是child、adult还是old。\end{tabular} & 在黑板上写字的老师一转身<CHILD> \\ \cmidrule(r){1-1} \cmidrule(l){3-4} 
\textbf{STTC} &  &  \begin{tabular}[c]{@{}c@{}}首先将语音转化为书面语，随后以<开始回答>\\作分隔，最后对语音内容做出答复。\end{tabular} & \begin{tabular}[c]{@{}c@{}}我感觉不太满意<开始回答>\\ 抱歉，我们会让你满意的。\end{tabular} \\ \bottomrule
\end{tabular}
\end{table}

\paragraph{Whisper Fine-tuning}
The original Whisper model supports a limited scope of speech-understanding tasks, which makes the direct integration of the Whisper with an LLM for multi-task training risky when data and computation resources are constrained.
Therefore, we first fine-tune the Whisper via multi-task data to ensure faster convergence of the OSUM model. 
Furthermore, this stage allows us to verify the reliability of our multi-task data. 
Specifically, we expand Whisper's instruction tag set to accommodate more tasks. Each forward pass executes only a single task. 

\paragraph{OSUM Training}
Training SULMs typically begins with pre-training on an ASR task, which serves as a foundation for incorporating additional speech tasks to enable LLMs to process semantic content from the speech encoder.
Given computational constraints, we introduce an ASR+X paradigm for OSUM's multi-task training.
It concurrently trains ASR and a secondary task ``X'', accelerating training while allowing the ``X'' task to utilize both textual and acoustic features, thereby potentially improving performance
The ASR+X paradigm follows a two-step process: first, transcribing speech to text (ASR); then, integrating this transcription with acoustic features to execute the target task (X). 
This is achieved within the LLM's autoregressive framework by adjusting predicted labels, without modifications to the model architectures or loss functions.
We implemented the ASR+X paradigm by prompting the LLLM with natural language prompts.
ChatGPT\footnote{\url{https://openai.com/index/chatgpt/}} is used to generate 5 candidate prompts for each task, one of which is randomly selected during training.
Table~\ref{tab:prompt_label} shows examples of the prompts and ASR+X prediction labels.

\subsection{Training Data}
\label{sec:training_data}
Our OSUM is designed to perform multi-task training using diverse speech datasets, with the goal of building a unified model capable of comprehensively understanding input speech in conversational scenarios.
The multi-task training process enables tasks to benefit from shared learning, enhancing overall model performance.
Upon completion of training, OSUM can be utilized for speech data annotation or further extended into a conversational Speech LLM. Detailed information about the datasets used for training is provided in Table~\ref{tab:train_data}.

\begin{table}[ht]
\centering
\footnotesize
\caption{Details of the multi-task training data for OSUM. The total duration is 50,500 hours.
The total training data includes both open-sourced datasets and the data we processed.
}
\label{tab:train_data}
\resizebox{\columnwidth}{!}{%
\begin{tabular}{@{}cccccc@{}}
\toprule
\textbf{Task} & \textbf{Total Hours} & \textbf{Language} & \textbf{Classification} & \textbf{Open-sourced Dataset} & \textbf{Processed Data} \\ \midrule
\textbf{ASR} & 24k & EN, CN & --- & \begin{tabular}[c]{@{}c@{}}Wenetspeech~\citep{wenetspeech}, \\ AISHELL-l~\citep{aishell1}, \\ AISHELL-2~\citep{aishell2}, \\ LibriSpeech~\citep{librispeech}\end{tabular} & No \\ \midrule
\textbf{SRWT} & 3k & EN, CN & --- & \begin{tabular}[c]{@{}c@{}}Wenetspeech~\citep{wenetspeech}, \\ AISHELL-1~\citep{aishell1}, \\ AISHELL-2~\citep{aishell2}, \\ LibriSpeech~\citep{librispeech}\end{tabular} & Yes \\ \midrule
\textbf{VED} & 2.5k & CN & \begin{tabular}[c]{@{}c@{}}Sign, Cry, Screaming, Laugh, Cough,\\ Throat Clearing, Sneeze, Other\end{tabular} & \begin{tabular}[c]{@{}c@{}}Audioset~\citep{jort_audioset_2017},\\ ESC-50~\citep{piczak2015dataset},\\ Vocal Sound~\citep{gong2022vocalsound},\\Nonspeech7k~\citep{rashid2023nonspeech7k},\\AISHELL-2~\citep{aishell2}\end{tabular} & Yes \\ \midrule
\textbf{SER} & 1k & EN, CN & \begin{tabular}[c]{@{}c@{}}Neutral, Happy, Anger, Sad, \\ Disgust, Fear, Surprise, Other\end{tabular} & \begin{tabular}[c]{@{}c@{}}ESD~\citep{zhou2021esd},\\ IEMOCAP~\citep{busso2008iemocap},\\ MSP-Podcast~\citep{martinez2020msp},\\ MER2023~\citep{Lian2023MER},\\ MELD~\citep{2019meld},\\ BIIC-Podcast~\citep{2023BIIC_Podcast}\end{tabular} & Yes \\ \midrule
\textbf{SSR} & 2.2k & CN & \begin{tabular}[c]{@{}c@{}}新闻科普，恐怖故事，童话故事，客服，\\ 诗歌散文，有声书，日常口语，其他\end{tabular} & Only internal datasets are used. & Yes \\ \midrule
\textbf{SGC} & 7.5k & EN, CN & Male, Female & \begin{tabular}[c]{@{}c@{}}KeSpeech~\citep{tang2021kespeech},\\  Datatang-Kid~\citep{datatang_page}, \\ Magicdata-Read~\citep{magicdata_read},\\ AISHELL-l~\citep{aishell1}, \\ AISHELL-2~\citep{aishell2}, \\LibriSpeech~\citep{librispeech},\\Kaggle-CommonVoice~\citep{kagglecv}\end{tabular} & No \\ \midrule
\textbf{SAP} & 6.8k & EN, CN & Child, Adult, Old & \begin{tabular}[c]{@{}c@{}}KeSpeech~\citep{tang2021kespeech},\\  Datatang-Kid~\citep{datatang_page}, \\ Magicdata-Read~\citep{magicdata_read}, \\ AISHELL-ASR0060~\citep{aishell_page}, \\ AISHELL-ASR0018~\citep{aishell_page},\\Kaggle-CommonVoice~\citep{kagglecv}\end{tabular} & No \\ \midrule
\textbf{STTC} & 3.5k & CN & --- & \begin{tabular}[c]{@{}c@{}}LCCC~\citep{2020lccc} \\ Wenetspeech~\citep{wenetspeech} \\ Databaker-Conversation~\citep{databaker}\end{tabular} & Yes \\ \bottomrule
\end{tabular}%
}
\end{table}

\subsection{Data Processing Pipe-line}
\label{sec:data_proccessing}
The data processing pipeline is crucial for training multi-task SULMs. In this section, we reveal the data processing schemes used for each task in the OSUM project, with the aim of providing a valuable reference for academic research.

\paragraph{ASR}
The training data include publicly available resources like Wenetspeech~\citep{wenetspeech}, AISHELL-1~\citep{aishell1}, AISHELL-2~\citep{aishell2}, and LibriSpeech~\citep{librispeech}, along with our internal ASR dataset, resulting in a total of 24,000 hours. 

\paragraph{SRWT}
For the SRWT task, a Gaussian Mixture Model - Hidden Markov Model (GMM-HMM) based conventional ASR model, is used to conduct force alignment and obtain word-level timestampes.
This model is trained on the 54,000-hour proprietary ASR dataset.
To evaluate its performance, we establish an internal SRWT test set and assess alignment quality using the Average Alignment Score (AAS) metric~\citep{shi2022aas}.
The GMM-HMM model achieves an AAS of 7.55, demonstrating its efficacy in generating reliable word-level timestamps.

\paragraph{SSR}
Given the absence of open-sourced tools for annotating style labels directly from audio data, we leverage two text-based LLMs-Qwen2.5-14B~\footnote{\url{https://huggingface.co/Qwen/Qwen2.5-14B-Instruct}} and GLM-4-9B-Chat~\footnote{\url{https://huggingface.co/THUDM/glm-4-9b-chat}}- to annotate speech transcriptions using carefully designed prompts.
To enhance annotation accuracy and reliability, we retain only the intersection of labeling results from both models. This intersection-based approach ensures high-quality annotations for training the SSR task.

\paragraph{VED}
We have attempted to train a vocal event labeling tool; however, due to the limited availability of training data, its classification performance is suboptimal, especially when vocal events and speech occur within the same audio segment. 
Therefore, we employ a Voice Conversion (VC) tool to modify the timbre of vocal event audio and insert it randomly into speech audio, creating a dataset of ASR+VED format. We find that this approach effectively mitigates the overfitting problems caused by the scarcity of vocal event training data with the assistance of VC. The open-source vocal event datasets we use include Audioset~\citep{jort_audioset_2017}, ESC-50~\citep{piczak2015dataset}, Vocal Sound~\citep{gong2022vocalsound}, and Nonspeech7k~\citep{rashid2023nonspeech7k}, while the ASR data consists solely of AISHELL-2~\citep{aishell2}.

\paragraph{SER}
Emotion2Vec~\citep{ma-etal-2024-emotion2vec} is the first universal speech emotion representation model. Without additional fine-tuning, we directly apply the pre-trained Emotion2Vec+ Large model~\footnote{\url{https://huggingface.co/emotion2vec/emotion2vec_plus_large}}, which is trained on 40,000 hours of emotional speech data, to annotate the audio with emotion labels.
Additionally, we leverage the GLM-4-9B-Chat model to generate emotion labels from the textual transcriptions of the speech.
By intersecting these annotations, we generate high-quality emotional labels for the entire dataset.

\paragraph{SGC}
Efforts to train a speaker gender classification model to label web-sourced data yield unsatisfactory performance.
Consequently, we discard the pseudo-labeled data and relied solely exclusively on manually labeled datasets for training.
For the SGC task, we select KeSpeech~\citep{tang2021kespeech}, Datatang-Kid~\citep{datatang_page}, AISHELL-1~\citep{aishell1}, AISHELL-2~\citep{aishell2}, LibriSpeech~\citep{librispeech}, Kaggle-CommonVoice~\citep{kagglecv}, and Magicdata-Read~\cite{magicdata_read} as training dataset, as they include reliable speaker gender labels. In addition, we employ a noisy data augmentation strategy.

\paragraph{SAP}
Similar to the SGC task, due to the poor performance of the automated labeling model, only manually labeled data is used for training.
We use KeSpeech~\citep{tang2021kespeech}, Datatang-Kid~\citep{datatang_page}, Magicdata-Read~\citep{magicdata_read}, Kaggle-CommonVoice~\citep{kagglecv}, AISHELL-ASR0060~\citep{aishell_page}, and AISHELL-ASR0018~\citep{aishell_page} as the training dataset for the SAP task, as these datasets provide reliable speaker age labels. Noisy data augmentation is also utilized in this task.

\paragraph{STTC}
For the STTC task, we use three types of data. First, we utilize a human-recorded audio question-answer dataset Databacker-Conversation~\citep{databaker}. Then, we use a text-based dialogue dataset LCCC~\citep{2020lccc} and the ChatTTS~\footnote{\url{https://github.com/2noise/ChatTTS}} system with random speaker capabilities to generate the utterances of the questioner in the dialogue, thus obtaining the speech-text pairs for the dialogue task. Finally, we filter suitable response sentences from the Wenetspeech~\citep{wenetspeech} dataset using Qwen2.5-7B~\footnote{\url{https://huggingface.co/Qwen/Qwen2.5-7B}}, guiding the LLM to generate text answers.

\section{Experiments}
\label{sec:experiments}
This section begins by presenting our training setup in Section~\ref{sec:exp_training_setup}. Subsequently, to conduct a more comprehensive evaluation of OSUM, we establish a series of complete internal evaluation sets, as detailed in Section~\ref{sec:exp_test_sets}. Finally, we report the performance of OSUM on both public and internal test sets, accompanied by an analysis in Section~\ref{sec:exp_main_res}.

\subsection{Training Setup}
\label{sec:exp_training_setup}
The two-stage training process for OSUM is detailed as follows:
\paragraph{Whisper Fine-tuning}
In the first stage, we fine-tune the Whisper-Medium model on the multi-task datasets described in Table~\ref{tab:train_data}.
This stage is conducted on 8 Nvidia A6000 GPUs.
A warm-up scheduler is employed to adjust the learning rate, peaking at 5e-5.
The multitask Whisper is trained for 150,000 steps, which takes approximately 15.3 days.

\paragraph{OSUM Training}
In the second stage, we conduct experiments on 24 Huawei Ascend NPUs, using a learning rate of 5e-5. The process completes a total of 704,000 training steps and consumes 10 days.

\begin{table}[h!]
\centering
\footnotesize
\caption{Details of the internal multi-task test sets. The categories in the classification task are balanced.}
\label{tab:test_set}
\resizebox{\columnwidth}{!}{%
\begin{tabular}{@{}cccc@{}}
\toprule
\textbf{Task} & \textbf{Name} & \textbf{Description} & \textbf{Item Number} \\ \midrule
\textbf{ASR} & $\text{Test}_\text{asr}^{1-5}$ & \begin{tabular}[c]{@{}c@{}}SpeechIO~\tablefootnote{\url{https://github.com/SpeechColab/Leaderboard}} is a widely used Chinese ASR leaderboard that includes \\
various scenarios such as TV programs, speeches, and news. \\
We use SpeechIO 0-4 as our ASR test set.\end{tabular} & 11932 \\ \midrule
\textbf{SRWT} & $\text{Test}_\text{srwt}$ & \begin{tabular}[c]{@{}c@{}}This test set features human-recorded audio from ten individuals, encompassing \\adult 
male and female voices as well as children's voices. It includes both \\
storytelling and reading styles, accompanied by meticulously \\annotated word-level timestamps.\end{tabular} & 494 \\ \midrule
\textbf{VED} & $\text{Test}_\text{ved}$ & \begin{tabular}[c]{@{}c@{}}The test set consists of internally human-recorded audio from ten \\
adult male and female speakers,
all reading in Chinese Mandarin, with the \\
insertion of speech events occurring at random positions.\end{tabular} & 400 \\ \midrule
\textbf{SER} & $\text{Test}_\text{ser}$ & \begin{tabular}[c]{@{}c@{}}Test data is selected from five publicly available emotional evaluation sets: \\
IEMOCAP~\citep{busso2008iemocap}, MER2023~\citep{Lian2023MER}, 
M3ED~\citep{M3ED}, \\
MSP-IMPROV~\citep{MSP-IMPROV}, and ESD~\citep{zhou2021esd}, including both \\ Chinese and English, 
encompassing eight types of emotions.\end{tabular} & 3885 \\ \midrule
\textbf{SSR} & $\text{Test}_\text{ssr}$ & \begin{tabular}[c]{@{}c@{}}This test set comprises two components: web-crawled data annotated \\
with style labels and highly expressive data intended for TTS training. \\
It encompasses eight speaking styles, predominantly in Chinese.\end{tabular} & 3000 \\ \midrule
\textbf{SGC} & $\text{Test}_\text{sgc}$ & \begin{tabular}[c]{@{}c@{}}The internal test set includes data from two age categories, with a female-to-male \\ratio of 2:3. All recordings are from human-recorded dialogue scenarios. \end{tabular} & 3000 \\ \midrule
\textbf{SAP} & $\text{Test}_\text{sap}$ & \begin{tabular}[c]{@{}c@{}}The internal test set includes three age categories: child, adult, and old, with an \\equal distribution of 1:1:1. All data consists of human-recorded dialogue scene data.\end{tabular} & 3000 \\ \midrule
\textbf{STTC} & $\text{Test}_\text{sttc}$ & \begin{tabular}[c]{@{}c@{}}This test set is generated from a Chinese text dialogue dataset. Specifically, \\we synthesize the question text into audio using the Cosyvoice2 TTS model, \\while the answers remain in text form. The text dialogue dataset used for the \\test set is derived from the same source as the training set.\end{tabular} & 200 \\ \bottomrule
\end{tabular}%
}
\end{table}

\subsection{Internal Test Sets}
\label{sec:exp_test_sets}
Currently, most SULMs evaluate multi-task performance using publicly available English datasets~\citep{chu2023qwen_audio,chu2024qwen2_audio,radford2023whisper,song2024touchasp}.
However, as OSUM training incorporates a substantial amount of Chinese data, we have developed a series of internal multi-task test sets tailored for Chinese~\footnote{We plan to make the internal test sets publicly available in the future}.
These complement the publicly available English test sets, creating a more comprehensive evaluation framework.
To support the ASR+X paradigm, we further enhance the test sets with speech transcripts.
However, ASR metrics are used solely for internal reference to assess model convergence and will not be publicly reported.
Table~\ref{tab:test_set} presents a description of our internal multi-task test sets.

\subsection{Main Results}
\label{sec:exp_main_res}
Table~\ref{tab:res_asr} and Table~\ref{tab:res_multi} show the experimental results of our OSUM across various tasks. The results reveal that our approach achieves performance that is comparable to, and in many cases superior to, speech understanding models such as Qwen-audio, Qwen2-audio, Whisper, and SenseVoice. Furthermore, in this section, we will highlight the performance disparities between our model and other comparable approaches, while providing a detailed analysis of the challenges SULMs face in these tasks. We hope that these experiences can provide useful references for researchers.

\begin{table}[htp]
\centering
\footnotesize
\caption{Evaluation results of ASR tasks on public and internal test sets.
The \textbf{bold} font represents the best result among the same test set. All internal results are inferred by ourselves.
}
\label{tab:res_asr}
\centering
\resizebox{0.9\columnwidth}{!}{%
\begin{tabular}{@{}ccccccc@{}}
\toprule
\textbf{Task} & \textbf{Model} & \textbf{Metric} & \textbf{\begin{tabular}[c]{@{}c@{}}Public \\ Test Set\end{tabular}} & \textbf{\begin{tabular}[c]{@{}c@{}}Public\\ Result\end{tabular}} & \textbf{\begin{tabular}[c]{@{}c@{}}Internal\\ Test Set\end{tabular}} & \textbf{\begin{tabular}[c]{@{}c@{}}Internal\\ Result\end{tabular}} \\ \midrule
\multirow{21}{*}{\textbf{ASR}} & Whisper-L-v3 & \multirow{21}{*}{\begin{tabular}[c]{@{}c@{}}WER/CER\\ (${\%}$, ${\downarrow}$)\end{tabular}} & \multirow{21}{*}{\begin{tabular}[c]{@{}c@{}}\textbf{WenetSpeech}\\ test-net/test-meeting\\ \textbf{AISHELL2}\\ mic/ios/android \\ \textbf{Librispeech}\\ test-clean/test-other\end{tabular}} & \begin{tabular}[c]{@{}c@{}}10.48/18.87\\ -/4.96/- \\ 1.82/3.50\end{tabular} & \multirow{21}{*}{\begin{tabular}[c]{@{}c@{}}$\text{Test}_\text{asr}^{1-3}$\\ $\text{Test}_\text{asr}^{4-5}$\end{tabular}} & \begin{tabular}[c]{@{}c@{}}11.91/5.24/8.22\\ 6.82/6.24\end{tabular} \\ \cmidrule(lr){2-2} \cmidrule(lr){5-5} \cmidrule(l){7-7} 
 & TouchASP &  &  & \begin{tabular}[c]{@{}c@{}}\textbf{5.52}/5.94\\ -/-/- \\ 2.03/4.38\end{tabular} &  & --- \\ \cmidrule(lr){2-2} \cmidrule(lr){5-5} \cmidrule(l){7-7} 
 & Qwen-Audio &  &  & \begin{tabular}[c]{@{}c@{}}-/-\\ 3.3/3.1/3.3 \\ 2.0/4.2\end{tabular} &  & \begin{tabular}[c]{@{}c@{}}4.84/1.18/4.98\\ 2.40/3.57\end{tabular} \\ \cmidrule(lr){2-2} \cmidrule(lr){5-5} \cmidrule(l){7-7} 
 & Qwen2-Audio &  &  & \begin{tabular}[c]{@{}c@{}}-/-\\ 3.0/3.0/2.9 \\ \textbf{1.6}/\textbf{3.6}\end{tabular} &  & \begin{tabular}[c]{@{}c@{}}2.70/1.08/4.34\\ \textbf{1.40}/3.10\end{tabular} \\ \cmidrule(lr){2-2} \cmidrule(lr){5-5} \cmidrule(l){7-7} 
 & Sensevoice-S &  &  & \begin{tabular}[c]{@{}c@{}}7.84/7.44\\ -/3.80/- \\3.15/7.18\end{tabular} &  & \begin{tabular}[c]{@{}c@{}}4.47/4.47/5.53\\ 5.21/5.16\end{tabular} \\ \cmidrule(lr){2-2} \cmidrule(lr){5-5} \cmidrule(l){7-7} 
 & Sensevoice-L &  &  & \begin{tabular}[c]{@{}c@{}}6.01/6.73\\ -/3.04/-\\2.57/4.28\end{tabular} &  & --- \\ \cmidrule(lr){2-2} \cmidrule(lr){5-5} \cmidrule(l){7-7} 
 & OSUM &  &  & \begin{tabular}[c]{@{}c@{}}6.46/\textbf{5.34} \\ \textbf{2.81}/\textbf{2.75}/\textbf{2.73} \\2.19/5.53\end{tabular} &  & \begin{tabular}[c]{@{}c@{}}\textbf{2.53}/\textbf{0.60}/\textbf{4.07}\\ 1.46/\textbf{3.04}\end{tabular} \\ \bottomrule
\end{tabular}%
}
\end{table}

\paragraph{ASR}
As illustrated in Table~\ref{tab:res_asr}, our approach reveals obvious advantages in the ASR task on the Chinese test sets.
Notably, the proposed OSUM consistently outperforms other models on the WenetSpeech test-meeting set, three AISHELL-2 sub-test sets, and four internally used SpeechIO test sets.
While OSUM does not surpass the top-performing method on the English test set, it rivals performance comparable to SenseVoice-S. These results are achieved, remarkably, with substantially less training data. 
In addition, we find that OSUM exhibits a surprisingly impressive ability to recognize Chinese-English code-mixed speech, even though such code-mixed datasets are not included during training. To be specific, the MER/CER/WER is ${3.89\%}$/${2.41\%}$/${19.30\%}$ on the ASRU code-switching test set~\citep{shi2020asru}.
Going forward, we will enhance this function. 
Overall, these results underscore that our ASR+X task paradigm effectively enhances model convergence in ASR tasks, significantly minimizing the data and computational resources required for SULMs training.

\paragraph{SRWT}
Table~\ref{tab:res_multi} presents the SRWT evaluation results for our proposed OSUM model compared to Whisper-Large-v3, Qwen-Audio, and the GMM-HMM model used for generating annotated data in SRWT tasks.
Our OSUM model significantly outperforms Whisper-Large-v3 by relative ${36.70\%}$ and also surpasses Qwen-Audio.
Additionally, our OSUM's performance in the SRWT task even slightly surpasses that of the GMM-HMM model, which is widely recognized for its high accuracy in timestamp prediction.
These results underscore the effectiveness of OSUM in the SRWT task.
Additionally, OSUM's high performance in the SRWT task not only enables it to predict timestamps in an end-to-end manner but, more importantly, simplifies its integration with other tasks, such as speaker diarization.

\paragraph{VED}
We first evaluate OSUM's performance on the public test sets ESC-50 and VocalSound. However, since the event categories in these two datasets do not completely align with those in OSUM, the comparison to other approaches should only serve as a rough assessment. 
Specifically, the ESC-50 contains a substantial number of non-vocal audio events, we categorize them as "other." The experimental results on this test set demonstrate that our model successfully classifies these non-vocal audio events as "other." Additionally, on the VocalSound set, we select the categories supported by OSUM and calculate the average accuracy across these categories.
This result reveals that our OSUM exhibits a gap compared to Qwen2-audio, primarily due to our training data consisting of concatenated speech and vocal events. In contrast, the VocalSound test set includes only the latter, resulting in a significant mismatch. Nevertheless, our OSUM achieves a norm level, successfully identifying the most independent vocal events. 
In our internal human-recoded ASR+VED test set, PANNs become unusable due to similar mismatches, particularly because their design treats speech as a standalone event, exacerbating accuracy degradation. Qwen2-audio performs relatively better but also experiences a performance decline in our test set, likely due to overfitting. In contrast, our model demonstrates balanced results in both the public and internal test sets, showcasing enhanced generalization. This indicates that using VC to augment data for vocal events can effectively mitigate overfitting in VED tasks.

\begin{table}[ht!]
\centering
\footnotesize
\caption{Evaluation results of multi-tasking on public and internal test sets. The best results for each test set are highlighted in \textbf{bold} font. 
Results shown in \textcolor{blue}{blue} font, as well as those on internal test sets, are inferred using the original released model by ourselves. 
}
\label{tab:res_multi}
\centering
\resizebox{0.9\columnwidth}{!}{%
\begin{tabular}{@{}ccccccc@{}}
\toprule
\textbf{Task} & \textbf{Model} & \textbf{Metric} & \textbf{\begin{tabular}[c]{@{}c@{}}Public \\ Test Set\end{tabular}} & \textbf{\begin{tabular}[c]{@{}c@{}}Public\\ Result\end{tabular}} & \textbf{\begin{tabular}[c]{@{}c@{}}Internal\\ Test Set\end{tabular}} & \textbf{\begin{tabular}[c]{@{}c@{}}Internal\\ Result\end{tabular}} \\ \midrule
\multirow{5}{*}{\textbf{SRWT}} & Whisper-L-v3 & \multirow{5}{*}{\begin{tabular}[c]{@{}c@{}}AAS(ms, ${\downarrow}$) \\ \citep{shi2022aas}\end{tabular}} & \multirow{5}{*}{\begin{tabular}[c]{@{}c@{}}The comparison plan test \\ set is not open source.\end{tabular}} & --- & \multirow{5}{*}{$\text{Test}_\text{srwt}$} & 11.09 \\ \cmidrule(lr){2-2} \cmidrule(lr){5-5} \cmidrule(l){7-7} 
 & Qwen-Audio &  &  & --- &  & 9.17 \\ \cmidrule(lr){2-2} \cmidrule(lr){5-5} \cmidrule(l){7-7} 
 & GMM-HMM &  &  & --- &  & 7.55 \\ \cmidrule(lr){2-2} \cmidrule(lr){5-5} \cmidrule(l){7-7} 
 & OSUM &  &  & --- &  & \textbf{7.02} \\ \midrule
\multirow{10}{*}{\textbf{VED}} & Qwen2-Audio & \multirow{10}{*}{\begin{tabular}[c]{@{}c@{}}ACC\\ (${\%}$, ${\uparrow}$)\end{tabular}} & \multirow{10}{*}{\begin{tabular}[c]{@{}c@{}}\textbf{ESC-50}\\ \textbf{VocalSound}\end{tabular}} & \begin{tabular}[c]{@{}c@{}}-\\ 93.3\end{tabular} & \multirow{10}{*}{$\text{Test}_\text{ved}$} & 33.25 \\ \cmidrule(lr){2-2} \cmidrule(lr){5-5} \cmidrule(l){7-7} 
 & TouchASP &  &  & \begin{tabular}[c]{@{}c@{}}85.7\\ -\end{tabular} &  & --- \\ \cmidrule(lr){2-2} \cmidrule(lr){5-5} \cmidrule(l){7-7} 
 & PANNs &  &  & \begin{tabular}[c]{@{}c@{}}83.3\\ -\end{tabular} &  & 3.25 \\ \cmidrule(lr){2-2} \cmidrule(lr){5-5} \cmidrule(l){7-7} 
 & OSUM &  &  & \begin{tabular}[c]{@{}c@{}}96.60\\ 82.58\end{tabular} &  & \textbf{78.25} \\ \midrule
\multirow{16}{*}{\textbf{SER}} & Qwen2-Audio & \multirow{16}{*}{\begin{tabular}[c]{@{}c@{}}ACC\\ (${\%}$, ${\uparrow}$)\end{tabular}} & \multirow{16}{*}{\begin{tabular}[c]{@{}c@{}}\textbf{MELD}\\ test\\ \textbf{MER2023}\\ test\end{tabular}} & \begin{tabular}[c]{@{}c@{}}55.3\\ -\end{tabular} & \multirow{16}{*}{$\text{Test}_\text{ser}$} & 38.04 \\ \cmidrule(lr){2-2} \cmidrule(lr){5-5} \cmidrule(l){7-7} 
 & TouchASP &  &  & \begin{tabular}[c]{@{}c@{}}50.5\\ -\end{tabular} &  & --- \\ \cmidrule(lr){2-2} \cmidrule(lr){5-5} \cmidrule(l){7-7} 
 & Sensevoice-L &  &  & \begin{tabular}[c]{@{}c@{}}\textbf{63.1}\\ 69.2\end{tabular} &  & --- \\ \cmidrule(lr){2-2} \cmidrule(lr){5-5} \cmidrule(l){7-7} 
 & Sensevoice-S &  &  & \begin{tabular}[c]{@{}c@{}}57.8\\ 68.3\end{tabular} &  & 40.77 \\ \cmidrule(lr){2-2} \cmidrule(lr){5-5} \cmidrule(l){7-7} 
 & Emotion2Vec &  &  & \begin{tabular}[c]{@{}c@{}}51.88\\ ---\end{tabular} &  & 51.14 \\ \cmidrule(lr){2-2} \cmidrule(lr){5-5} \cmidrule(l){7-7}
 & EmoBox &  &  & \begin{tabular}[c]{@{}c@{}}51.89\\ 65.23\end{tabular} &  & \textbf{74.54} \\ \cmidrule(lr){2-2} \cmidrule(lr){5-5} \cmidrule(l){7-7} 
 & OSUM &  &  & \begin{tabular}[c]{@{}c@{}}53.38\\ \textbf{86.43}\end{tabular} &  & 72.33 \\ \midrule
\multirow{2}{*}{\textbf{SSR}} & GLM-4 & \multirow{2}{*}{\begin{tabular}[c]{@{}c@{}}ACC\\ (${\%}$, ${\uparrow}$)\end{tabular}} & \multirow{2}{*}{\begin{tabular}[c]{@{}c@{}}The comparison plan test\\ set is not open source.\end{tabular}} & --- & \multirow{2}{*}{$\text{Test}_\text{ssr}$} & 53.97 \\ \cmidrule(lr){2-2} \cmidrule(lr){5-5} \cmidrule(l){7-7} 
 & OSUM &  &  & --- &  & \textbf{67.34} \\ \midrule
\multirow{4}{*}{\textbf{SGC}} & Qwen2-Audio & \multirow{4}{*}{\begin{tabular}[c]{@{}c@{}}ACC\\ (${\%}$, ${\uparrow}$)\end{tabular}} & \multirow{4}{*}{\begin{tabular}[c]{@{}c@{}}\textbf{AISHELL-1}\\ test\\ \textbf{Kaggle-CommonVoice}\\ valid-test\end{tabular}} & \begin{tabular}[c]{@{}c@{}}\textcolor{blue}{97.36}\\ \textcolor{blue}{97.25}\end{tabular} & \multirow{4}{*}{$\text{Test}_\text{sgc}$} & \textbf{98.43} \\ \cmidrule(lr){2-2} \cmidrule(lr){5-5} \cmidrule(l){7-7} 
 & OSUM &  &  & \begin{tabular}[c]{@{}c@{}}\textbf{100}\\ \textbf{99.41}\end{tabular} &  & 96.79 \\ \midrule
\multirow{2}{*}{\textbf{SAP}} & Qwen2-Audio & \multirow{2}{*}{\begin{tabular}[c]{@{}c@{}}ACC\\ (${\%}$, ${\uparrow}$)\end{tabular}} & \multirow{2}{*}{\begin{tabular}[c]{@{}c@{}}\textbf{Kaggle-CommonVoice}\\ valid-test\end{tabular}} & \textcolor{blue}{35.53} & \multirow{2}{*}{$\text{Test}_\text{sap}$} & 49.52 \\ \cmidrule(lr){2-2} \cmidrule(lr){5-5} \cmidrule(l){7-7} 
 & OSUM &  &  & \textbf{76.52} &  & \textbf{67.36} \\ \midrule
\multirow{2}{*}{\textbf{STTC}} & Qwen2-Audio & \multirow{2}{*}{\begin{tabular}[c]{@{}c@{}}GPT-3.5-Turbo\\ Scoring\end{tabular}} & \multirow{2}{*}{\begin{tabular}[c]{@{}c@{}}\textbf{AirBench}\\ speech\end{tabular}} & \textcolor{blue}{6.77} & \multirow{2}{*}{$\text{Test}_\text{sttc}$} & 6.91 \\ \cmidrule(lr){2-2} \cmidrule(lr){5-5} \cmidrule(l){7-7} 
 & ASLP-Audio &  &  & 5.98 &  & \textbf{7.04} \\ \bottomrule
\end{tabular}%
}
\end{table}

\paragraph{SER}
For the SER task, we extract the categories supported by OSUM from the public datasets MELD and MER2023 for testing, followed by a comprehensive evaluation on our internal test set.
In the experiments with the public datasets, OSUM demonstrates superior performance on the MER2023 test set, outperforming several recent public benchmark models. On the MELD dataset, OSUM's performance ranks just below the SenseVoice-L model, likely due to the latter's additional training on a larger-scale speech emotion dataset. In addition, while OSUM's result on the internal test set is comparable to that of the EmoBox model, it significantly surpasses other comparative approaches. Furthermore, we observe that among the eight emotions supported, disgust and fear are particularly challenging to recognize, a difficulty partly attributed to the scarcity of training data for these two emotions. In our future work, we plan to enhance the model's performance and generalization capability by utilizing OSUM for labeling, thereby obtaining a larger and more balanced emotion dataset.

\paragraph{SSR}
The acoustic-text dual-modal style classification employed by our OSUM significantly outperforms the single-text modality of GLM-4-9B-Chat.
It demonstrates a strong ability to distinguish among eight styles: news and science reporting, horror stories, fairy tales, customer service, poetry and prose, audiobooks, spontaneous conversation, and others.
Notably, the classification performance for news science communication, audiobooks, fairy tales, and customer service styles is commendable; however, there remains room for improvement in the categorization of poetry and prose, horror stories, and other styles. Moving forward, we leverage OSUM to label additional data, aiming to enhance data quality and optimize the distribution across categories.

\paragraph{SGC}
In the SGC task, we evaluate Qwen2-Audio and OSUM. 
The results demonstrate that OSUM achieves an ${100\%}$ accuracy on the AISHELL-1 test set. While this result is commendable, we suspect it may indicate some degree of overfitting. Furthermore, on the Kaggle test set, our approach slightly outperforms Qwen2-Audio, yet it falls short on our internal test set. 
This indicates that Qwen2-Audio has strong robustness in the SGC task, which may be due to the fact that they have used a wider range of data. In the follow-up, we plan to continue to increase the training data for this task. 
Overall, OSUM exhibits its value in the SGC task.

\paragraph{SAP}
We also compare our OSUM with Qwen2-Audio on the SAP task.
During our previous experiments, we found that the acoustic similarity between teenagers and adults is remarkably high, complicating effective differentiation.
Consequently, we categorize age into three groups: child, adult, and old. 
Curiously, despite our efforts to debug the prompts, Qwen2-Audio demonstrates a lower age classification accuracy on both the Kaggle test set and our internal test set. This may stem from their overly detailed age categorization, which hinders the model's training accuracy. Our model significantly surpasses Qwen2-Audio on the Kaggle test set, achieving an accuracy of ${76.52\%}$. Although the classification accuracy slightly declines on our proprietary test set, it still outperforms Qwen2-Audio. This indicates that our model exhibits strong generalization capabilities on different data.

\paragraph{STTC}
In the STTC task, we follow AirBench's evaluation protocol across all test sets. This involves providing the text of audio queries along with the text of two distinct answers, allowing a text-based LLM to assign subjective scores from 1 to 10. The two answers consist of a real response and the answer generated by SULMs. While AirBench employs GPT-4 as the scoring LLM, it is currently inaccessible, so we instead utilize GPT-3.5-Turbo.
The test results presented in Table~\ref{tab:res_multi} indicate that, on AirBench's official speech sub-test set, our score is lower than that of Qwen2-Audio, suggesting that our model's capabilities in English conversation and audio description lag behind those of Qwen2-Audio. 
This is primarily because we do not use English conversational data for training; the current score relies entirely on the LLM's performance.
However, on our internal Chinese conversational test set, OSUM outperforms Qwen2-Audio, which indicates that our strategy of performing ASR before chat is beneficial. Overall, our OSUM model is comparable to Qwen2-Audio in terms of conversational ability. We will not be content with this achievement. Our future work will mainly focus on the conversation task.  
\section{Future Works}
\label{sec:future_work}

While OSUM demonstrates commendable performance, our research remains an ongoing endeavor to push the boundaries of academic exploration in Speech LLMs. In the coming months, we aim to address several key areas for improvement and innovation:

\begin{itemize}

    \item \textbf{Expanding OSUM’s Functionalities.} We plan to enhance OSUM with additional capabilities, such as language and accent identification, to broaden its applicability in multilingual and diverse speech scenarios.

    \item \textbf{Enabling Multi-Task Capability.} We aim to activate OSUM's ability to perform multiple tasks simultaneously, such as identifying the emotion, age, gender, and speaking style in a single inference pass. Leveraging this multi-task capability, we plan to develop a versatile data labeling tool to streamline audio data processing pipelines.

    \item \textbf{Incorporating Full-Duplex Voice Interaction.} To improve naturalness and responsiveness, we plan to integrate full-duplex voice interaction capabilities into OSUM. This enhancement will allow OSUM to generate context-aware, natural responses, such as matching the questioner’s emotion or mimicking specific speaking styles, like that of a child.

    \item \textbf{Open Science Contributions.} As part of our commitment to advancing academic research, we will continue to share detailed training methodologies, data pipelines, and model updates. Our aim is to foster collaboration, provide valuable resources for researchers, and democratize access to cutting-edge Speech LLM technologies.
    
\end{itemize}

Through these efforts, we seek to extend OSUM’s capabilities, establish new benchmarks for Speech LLMs, and contribute meaningfully to the academic study and practical applications of speech understanding.
\section{Conclusion}
\label{sec:conclusion}
In this study, we propose OSUM, an open-source Speech understanding language model. The OSUM model integrates a Whisper encoder with a Qwen2 LLM, supporting eight speech tasks. By employing an ASR+X training strategy, OSUM achieves efficient and stable multi-task training, simultaneously optimizing ASR alongside target tasks. Beyond delivering robust performance, OSUM prioritizes transparency by providing openly accessible data preparation and training methodologies, offering valuable insights and practical guidance for the academic community.

\section{Acknowledgment}
\label{sec:acknowledgment}
We appreciate that AISHELL~\footnote{\url{https://www.aishelltech.com}}, Databaker~\footnote{\url{https://www.data-baker.com}}, MagicData~\footnote{\url{https://www.magicdatatech.com}}, and DataTang~\footnote{\url{https://www.datatang.com}} generously provide some training data.
Most of our model components are trained using Huawei~\footnote{\url{https://www.huawei.com}} Ascend NPUs.

\section{Authors}
\label{sec:authors}
The ranking is from top to bottom and then from left to right. $^{\ast}$ indicates equal contribution. ~\textsuperscript{\dag} indicates the corresponding author.

\begin{multicols}{3}
    \begin{itemize}[noitemsep]
        \item Xuelong Geng  
        \item Kun Wei 
        \item Qijie Shao 
        \item Shuiyun Liu$^{\ast}$ 
        \item Zhennan Lin$^{\ast}$ 
        \item Zhixian Zhao$^{\ast}$ 
        \item Guojian Li$^{\ast}$ 
        \item Wenjie Tian$^{\ast}$ 
        \item Peikun Chen 
        \item Yangze Li 
        \item Pengcheng Guo 
        \item Mingchen Shao 
        \item Shuiyuan Wang 
        \item Yuang Cao 
        \item Chengyou Wang 
        \item Tianyi Xu 
        \item Yuhang Dai 
        \item Xinfa Zhu 
        \item Yue Li 
        \item Li Zhang 
        \item Lei Xie~\textsuperscript{\dag} 
    \end{itemize}
\end{multicols}

\vfill\pagebreak

\bibliographystyle{plainnat} 
\bibliography{8_Ref}

\end{CJK}
\end{document}